\documentclass[sigconf,nonacm]{acmart}
\AtBeginDocument{%
  \providecommand\BibTeX{{%
    \normalfont B\kern-0.5em{\scshape i\kern-0.25em b}\kern-0.8em\TeX}}}

\setcopyright{acmcopyright}
\copyrightyear{2023}
\acmYear{2023}
\acmDOI{XXXXXXX.XXXXXXX}

\acmConference['JCDL]{ACM/IEEE Joint Conference on Digital Libraries}{June 26--30,
  2018}{Santa Fe, NM}
\acmPrice{15.00}
\acmISBN{978-1-4503-XXXX-X/18/06}

\begin{document}

%%
%% The "title" command has an optional parameter,
%% allowing the author to define a "short title" to be used in page headers.
\title{Maximizing Equitable Reach and Accessibility of ETDs}

%%
%% The "author" command and its associated commands are used to define
%% the authors and their affiliations.
%% Of note is the shared affiliation of the first two authors, and the
%% "authornote" and "authornotemark" commands
%% used to denote shared contribution to the research.
\author{William A. Ingram}
\email{waingram@vt.edu}
\orcid{0000-0002-8307-8844}
\affiliation{%
  \institution{Virginia Tech}
  \streetaddress{560 Drillfield Drive}
  \city{Blacksburg}
  \state{Virginia}
  \country{USA}
  \postcode{24061}
}

\author{Jian Wu}
\email{j1wu@odu.edu}
\orcid{0000-0003-0173-4463}
\affiliation{%
  \institution{Old Dominion University}
  \city{Norfolk}
  \state{Virginia}
  \country{USA}
}
\author{Edward A. Fox}
\email{fox@vt.edu}
\orcid{0000-0003-1447-6870}
\affiliation{%
  \institution{Virginia Tech}
  \streetaddress{}
  \city{Blacksburg}
  \state{Virginia}
  \country{USA}
}

\begin{abstract}
This poster addresses accessibility issues of electronic theses and dissertations (ETDs) in digital libraries (DLs). 
ETDs are available primarily as PDF files, which present barriers to equitable access, especially for users with visual impairments, cognitive or learning disabilities, or for anyone needing more efficient and effective ways of finding relevant information within these long documents.
We propose using AI techniques, including natural language processing (NLP), computer vision, and text analysis, to convert PDFs into machine-readable HTML documents with semantic tags and structure, extracting figures and tables, and generating summaries and keywords. 
Our goal is to increase the accessibility of ETDs and to make this important scholarship available to a wider audience. 
% Our work demonstrates the potential of AI to significantly improve the accessibility of ETDs and highlights the importance of making this valuable research accessible to all.
\end{abstract}

%%
%% The code below is generated by the tool at http://dl.acm.org/ccs.cfm.
%% Please copy and paste the code instead of the example below.
%%


%%
%% Keywords. The author(s) should pick words that accurately describe
%% the work being presented. Separate the keywords with commas.

%% A "teaser" image appears between the author and affiliation
%% information and the body of the document, and typically spans the
%% page.
% \begin{teaserfigure}
%   \includegraphics[width=\textwidth]{sampleteaser}
%   \caption{Seattle Mariners at Spring Training, 2010.}
%   \Description{Enjoying the baseball game from the third-base
%   seats. Ichiro Suzuki preparing to bat.}
%   \label{fig:teaser}
% \end{teaserfigure}

% \received{20 February 2007}
% \received[revised]{12 March 2009}
% \received[accepted]{5 June 2009}

%%
%% This command processes the author and affiliation and title
%% information and builds the first part of the formatted document.
\maketitle

\section{Introduction}
University-based institutional repositories are DL systems used to manage, preserve, and distribute intellectual output from faculty, staff, and students.
They often contain a significant number of ETDs, the final product of graduate students' research, which are typically long, book-length documents.
The most common format for ETDs is the Portable Document Format (PDF), which is widely used as it preserves the visual formatting and layout of the document and is compatible with most computer systems. 
% ETDs make a significant contribution to the existing body of knowledge in their field.
% They often include a detailed literature review, references, figures,  tables, data analysis, and novel research findings. 
PDFs have many advantages for scholarly work, but their lack of machine readability and broad accessibility through assistive devices is a significant limitation. 

% Nearly since their inception, ETDs have been tied to the use of PDF. 
The first ETDs were created around 1988 as Standard Generalized Markup Language (SGML) documents. 
However, widespread adoption of ETDs did not occur until the introduction of PDF and the release of Adobe's Acrobat tool in the early 1990s. 
Before the release of the first version of PDF and Adobe Acrobat in 1993, the ETD team at Virginia Tech, through a partnership with Adobe, was able to evaluate a pre-release version of the software to explore its potential for ETDs~\cite{fox_electronic_2012}.
Their efforts helped lay the foundation for ETDs and aided the widespread adoption of PDF for the dissemination of scholarly work.
ETDs are often only available as PDFs, which typically lack machine readability, semantic structure, and navigation elements, making it difficult to interact with the content, particularly for users with visual impairments or other disabilities. 
As book-length documents, ETDs present unique barriers to access due to their length and complexity. 

% ETDs are still typically stored and distributed as PDFs, and while basic metadata is usually provided along with the PDF, the options for interacting with the content within the digital library itself are limited. 
% PDFs are not machine readable, and they often lack meaningful structure and navigation elements, such as headings, lists, and alt text for images. 
% Other formats, such as HTML or e-pub, provide more accessibility features for ETDs. 
% As book-length documents, ETDs present unique barriers to access compared to shorter scholarly works, such as journal articles and conference papers. 
% The length and complexity of an ETD can make it more challenging to navigate, discovering, and accessing the knowledge contained within book-length PDF documents is a significant usability problem, and it is particularly challenging for users with visual impairments or other disabilities, who may need additional accessibility features to find information within the document. 

% More advanced computational techniques are needed to make ETDs more accessible and discoverable to a wider audience.
There is a growing trend toward making scholarly works more machine readable and accessible through the use of tools such as PDF-to-HTML conversion, summarization, and keyword extraction. 
Advances in machine learning, deep learning, and NLP can improve the accessibility of ETDs, and increase and broaden their usefulness.

\section{Related Work}
% Iris Xie is widely acclaimed for her work on the usability experience for digital libraries (e.g.,~\cite{DBLP:journals/ipm/Xie08,iris_xie_evaluation_2006}). 
% Xie and her colleagues identified numerous usability problems that blind and visually impaired users experience that prevent them from effectively interacting with digital library content~\cite{DBLP:journals/ipm/XieBLCYH20}. 
% This research led to the development of the Digital Library Accessibility and Usability Guidelines (DLAUG) to Support Blind and Visually Impaired Users~\cite{xie_dlaug_2021} in 2021.
% The guidelines describe some of the problems blind and visually impaired users face when accessing PDFs, especially scanned PDFs, and suggest several techniques for making PDF files more accessible, such as inserting PDF tags and using optical character recognition software for scanned documents. 
% They also recommend providing the user with document summaries, keywords, and relevant document snippets. 
% Adherence to these guidelines is time-consuming and typically involves a lot of manual work by the authors. 
Iris Xie and her collaborators have written extensively on the usability of DLs (e.g.,~\cite{DBLP:journals/ipm/Xie08,iris_xie_evaluation_2006}). 
Her recent focus on the needs of blind and visually impaired users~\cite{DBLP:journals/ipm/XieBLCYH20} led to the development of the Digital Library Accessibility and Usability Guidelines (DLAUG) in 2021~\cite{xie_dlaug_2021}.
Some of the guidelines address problems with accessing PDFs, particularly scanned PDFs, and recommend several techniques to make PDF files more accessible to blind and visually impaired users. These include inserting PDF tags and using OCR software for scanned documents. 
The guidelines also recommend providing users with document summaries, keywords, and relevant document snippets. 
However, adherence to these guidelines is time consuming and typically involves manual work by the authors.

Usability advocate Jakob Nielsen has written for more than 25 years about problems PDF files cause online readers~\cite{nielsen_defense_1996,nielsen_experience_pdf_2020}. 
For long documents, Nielsen recommends generating two versions: one optimized for online viewing (HTML) and one optimized for printing (PDF)---but urges that PDF files should \textit{never} be read online~\cite{nielsen_defense_1996}. 
Nielsen advises designers to avoid PDFs unless a printable PDF is necessary. 
In these cases, he suggests creating a gateway page that summarizes key components and critical information from the document with the option to download the full PDF~\cite{nielsen_experience_pdf_2020}.

A framework for improving the accessibility of articles submitted to the arXiv.org preprint repository was recently published in 2022~\cite{DBLP:journals/corr/abs-2212-07286}.
The paper proposes that arXiv should offer an HTML version alongside the PDF and TeX formats currently offered. 
According to the article, 90\% of the submissions to arXiv are provided as TeX, but the conversion from TeX to HTML cannot be fully automated. 
Authors will need to adjust their workflows in order to create properly formatted HTML versions of their papers. 
Many efforts are being made to overcome the limitations of scholarly PDFs through the use of AI.
%The SciA11y project developed by AllenAI 
AllenAI's SciA11y project
aims to increase the accessibility of scientific documents by using AI and NLP techniques to extract and convert the semantic content of scientific PDFs into accessible HTML~\cite{DBLP:journals/corr/abs-2105-00076}. 
Our work is closely related. 
However, while it is possible that their system could be applied to ETDs, the focus of their work is on improving access to scientific papers (e.g., for conferences and journals), which are shorter and structured differently than theses and dissertations. 

\section{Preliminary Work}
Our team compiled a research corpus of more than 500,000 full text ETDs and metadata collected from 40+ institutional repositories of universities throughout the United States~\cite{ingram:uddin2021}. 
The corpus is widely diverse in terms of the departments and academic disciplines it represents.
By training models on a diverse corpus, we expose them to a wider range of writing styles, subject matter, and discourse conventions. 
Our aim is to increase the generalizability and adaptability of our models, making them better suited for working with a variety of ETDs from different fields and disciplines. 
Additionally, the inclusion of ETDs from multiple disciplines can help identify commonalities in the structure and content of ETDs in general, which could further improve the performance of our models.
In multiple studies, we trained models for various tasks with the goal of improving accessibility. 
These tasks include metadata extraction~\cite{ingram:choudhury2020,ingram:choudhury2021}, figure and table extraction~\cite{ingram:kahu2021}, summarization~\cite{ingram:ingram2020summarizing}, keyword generation~\cite{jude_increasing_2020}, topic modeling~\cite{ingram:ahujaetd2022}, and PDF-to-XML conversion~\cite{ahuja-etal-2022-parsing}.
By converting the PDF to XML, we capture the semantic structure of the document. 
The XML is converted to HTML or ePub for humans to read online, and it can be easily converted to other XML formats, such as the JATS format used by PubMed and others, to increase the interoperability and discoverability of the ETD, and allow for more efficient indexing, searching, and retrieval of content by other systems. 
By combining these techniques, we aim to create a more accessible, navigable, and machine-readable DL for ETDs. 

\balance

\section{Discussion and Future Work}
We investigate using AI to convert PDF ETDs to machine-readable HTML documents with semantic tags, extracted figures and tables, and generated summaries and keywords, with the aim of making them machine-readable and more accessible to a wider audience. 
More research is underway to assess the impact of the proposed techniques on the accessibility and usability of ETDs through user studies involving a diverse group of participants, including those with visual impairments and cognitive or learning disabilities. 

As ETDs are complex book-length documents, creating one long HTML representation might not be the best way to present them. 
More research is needed to determine how users can navigate and consume information in an ETD in the most effective and efficient way. 
ETDs differ from other academic writing in their length and format and contain a diverse range of content, including text, images, tables, equations, and references. 
A combination of approaches, including structured navigation, adaptive interfaces, and summarization, may be needed to support users in finding and understanding the content buried in these rich scholarly documents.
%%
%% The acknowledgments section is defined using the "acks" environment
%% (and NOT an unnumbered section). This ensures the proper
%% identification of the section in the article metadata, and the
%% consistent spelling of the heading.
\begin{acks}
This project was made possible in part by the \grantsponsor{imls}{Institute of Museum and Library Services}{https://www.imls.gov/} \grantnum{imls}{LG-37-19-0078-19}.
\end{acks}

%%
%% The next two lines define the bibliography style to be used, and
%% the bibliography file.
\bibliographystyle{ACM-Reference-Format}
\bibliography{base}

\end{document}